\documentclass{emulateapj}
\usepackage{graphicx,amsfonts,natbib,apjfonts}
\shorttitle{Dendrograms}
\shortauthors{Rosolowsky et al.}
\begin{document}
\title{Structural Analysis of Molecular Clouds: Dendrograms}
\author{E. W. Rosolowsky\altaffilmark{1,2},
  J. E. Pineda\altaffilmark{1}, J. Kauffmann\altaffilmark{1,3},  
  and A. A. Goodman\altaffilmark{1,3}} 
 \email{erosolow@cfa.harvard.edu}

\altaffiltext{1}{Center for Astrophysics, 60 Garden St., Cambridge,
  MA 02138}

\altaffiltext{2}{University of British Columbia Okanagan, 3333
  University Way, Kelowna, BC V1V 1V7, Canada}

\altaffiltext{3}{Initiative for Innovative Computing, Harvard
  University, 60 Oxford St., Cambridge, MA 02138}

\begin{abstract}
We demonstrate the utility of dendrograms at representing the
essential features of the hierarchical structure of the isosurfaces
for molecular line data cubes.  The dendrogram of a data cube is an
abstraction of the changing topology of the isosurfaces as a function
of contour level.  The ability to track hierarchical structure over a
range of scales makes this analysis philosophically different from
local segmentation algorithms like CLUMPFIND.  Points in the
dendrogram structure correspond to specific volumes in data cubes
defined by their bounding isosurfaces.  We further refine the
technique by measuring the properties associated with each isosurface
in the analysis allowing for a multiscale calculation of molecular gas
properties.  Using COMPLETE $^{13}$CO~$(J=1\to 0)$ data from the L1448
region in Perseus and mock observations of a simulated data cube, we
identify regions that have a significant contribution by self-gravity
to their energetics on a range of scales.  We find evidence for
self-gravitation on all spatial scales in L1448 though not in all
regions.  In the simulated observations, nearly all of the emission is
found in objects that would be self-gravitating if gravity were
included in the simulation.  We reconstruct the size-line width
relationship within the data cube using the dendrogram-derived
properties and find it follows the standard relation: $\sigma_v
\propto R^{0.58}$.  Finally, we show that constructing the dendrogram
of CO $(J=1\to 0)$ emission from the Orion-Monoceros region allows for
the identification of giant molecular clouds in a blended molecular
line data set using only a physically motivated definition
(self-gravitating clouds with masses $>5\times 10^4~M_{\odot}$).
\end{abstract}


\section{Introduction}

The structure in molecular clouds determines, in part, the locations,
numbers and masses of newly formed stars.  Because of its important
role at establishing the initial mass function of stars as well as the
local star formation rate, great effort has been invested in
characterizing the structure of this gas.  Observations of molecular
clouds including molecular and atomic line surveys, extinction and
infrared emission mapping, and star counts have all been used to
characterize the nature of molecular clouds.  A myriad of analytic
techniques have been applied to these data with a broad range of
results.  Each technique is designed to highlight a different feature
of the gas: fractal analysis techniques are used to
demonstrate that the gas is fractal \citep{stutzki98}; searches for
clumps utilize clump identification algorithms \citep{gaussclumps};
studies characterizing turbulence frequently aim to measure
theoretically relevant quantities such as the power spectrum
\citep{vca} or the structure function \citep{heyer-turb}.

One of the dominant characteristics of molecular gas it that it is
{\it hierarchical}.  A preponderance of multi-tracer studies have
consistently shown that the high-density features in molecular clouds
have relatively small physical scales and are invariably contained
inside envelopes of lower density gas \citep[e.g.][]{bs86,
  lada-dense}.  Moreover, the hierarchy is non-trivial: for any given
scale, there are more small-scale, dense structures than there are
large-scale, sparse structures.  Dense cores are the top level of the
cloud hierarchy and the turbulence that characterizes molecular clouds
makes a transition to coherence (i.e. domination by thermal rather
than turbulent motions) on $\ell\sim 0.1$~pc scales
\citep{goodman-coherence,tafalla04,lada-thermal}.  These dense cores are the
exclusive hosts of star formation inside molecular clouds and much
effort has been expended to study the properties of these cores and
the stars that form inside them \citep[e.g.,][and references
  therein]{pp5-difran,pp5-wt}.  Indeed, studies have argued for the
close relationship between the dense cores and the newly-formed stars
based on the similarities of their mass functions
\citep{motte-andre,testi-sargent,alves-pipe,bolocam-perseus}.

The low-density gas that fills the majority of the volume of the
molecular cloud can be regarded as the bottom of the gas hierarchy in
the molecular clouds (though the filling fraction and chemical state
of the molecular cloud is far from uniform).  The chemical change
associated with the formation of star-forming (molecular) clouds has
commonly been used to define discrete clouds in the interstellar
medium (ISM) serving as a useful division between the diffuse,
multi-phase ISM and star-forming clouds.  However, there is some
debate about whether the boundaries of molecular clouds form a
meaningful bottom of the hierarchy and are distinct entities
\citep[the ``classical'' interpretation,][]{psp5}; or whether the
hierarchical structure continues with only chemical changes into the
diffuse ISM \citep[e.g.][]{bvs99,hbb01}.  The crux of the debate
centers around lifetimes of the molecular clouds relative to their
internal crossing times or, equivalently, the importance of
self-gravity in the cloud's energetics. However, much of this debate
has centered on considering disparate sets of observations and
\citet{eg-rapid} has presented a synthesis that argues relatively
long-lived (20-30 Myr) self-gravitating clouds can accommodate local,
rapid star formation within them accounting for the sets of
observations that drive an apparent contradiction in measurements of
cloud lifetimes.

Connecting molecular clouds to the atomic gas in and around them is
particularly difficult since the 21-cm observations that would be
directly comparable to molecular line studies suffer from fore- and
background confusion as well as an intrinsic degradation of spatial
resolution from the long wavelength of the emission.  For some cases,
where geometry \citep{pound-uma}, self-absorption \citep{hinsa}, or
modeling of photodissociation regions \citep{bensch06} allows, the
atomic gas related to molecular clouds can be studied.  Studying the
hierarchical structure within molecular clouds requires a large
spatial dynamic range which restricts useful observational data sets
to galactic objects.  Although the hierarchical structure of the ISM
continues to large scales in the galaxies, the above restrictions
limit considerations of hierarchical structure within star forming
clouds to those found in the gas traced by molecular emission.

This paper presents another analytic technique aimed to characterize
the hierarchical structure in molecular gas and relate it to the star
formation process.  We use {\it dendrograms} to graphically represent
hierarchical structure of nested isosurfaces in three-dimensional
molecular line data cubes (i.e.~position-position-velocity data
cubes).  The dendrograms are abstractions of how the isosurfaces nest
inside one another.  Our principal contribution in this work is using
standard molecular line analysis techniques to characterize the
branches in a dendrogram allowing for simultaneous measurement of
various properties on a range of physical scales.  In addition,
dendrograms are a reduction of the structure in a data set to its
essential features and, as such, they become useful reductions of
large data sets to simple models allowing the study of a wide range of
spatial scales.

The dendrograms presented here are simply an alternative application
of the {\it structure trees} presented first in \citet[][ hereafter
  HS92]{houlahan2}.  While novel at the time for the star formation
community, such diagram techniques were relatively common in other
disciplines \citep{graphs}.  In the intervening time since the
publication of HS92, the analysis of tree networks has become even
more developed and tools for the construction and analysis of the
resulting structure trees have become commonplace (e.g. we will apply
software in the standard IDL distribution for the following analysis).
Our application of the dendrogram formalism uses a significantly
different analytic approach compared to the work of HS92.  They
analyzed the characteristics of the structure trees derived from
two-dimensional data.  The present work uses dendrograms as an
abstraction of the isosurfaces present in three-dimensional data,
emphasizing the properties of those isosurfaces.  Finally, note that
the application of dendrograms to contour surfaces as in this work and
HS92 is significantly different from their common application in
statistical analysis \citep[e.g.][]{ghaz99} where they are used to
represent clustering in statistical data sets.  We refer to HS92's
structure trees as {\it dendrograms} to be consistent with the
nomenclature adopted in other fields, in particular that of the
statistical description of hierarchical systems.

This paper briefly discusses different approaches to molecular line
data (\S\ref{mol-analysis}) before developing the concept of
dendrograms (\S\ref{dendrograms}).  We discuss several refinements of
the dendrogram technique including accounting for the effects of noise
(\S\ref{pruning}), measuring cloud properties on dendrogram branches
(\S\ref{cloudprops}), and the complications of mapping between
observed and physical domains (\S\ref{interp}).  We conclude with two
applications of the dendrogram technique: an analysis of self-gravity
in L1448 (\S\ref{l1448}) and the identification of GMCs in blended
data sets (\S\ref{gmcs}).

\section{The Analysis of Molecular Line Data}
\label{mol-analysis}
Broadly speaking, the statistical analysis of molecular line data has
usually followed one of two paths.  Either authors construct
statistical descriptions of the emission from an entire molecular line
data set, or authors will segment (divide) the data into what they
believe to be physically relevant structures and study the
distribution of properties in the resulting population of objects.
Common examples of statistical analysis include fractal analysis
\citep{fractal-mspec,stutzki98}, $\Delta$-variance
\citep{stutzki98,deltavar}, correlation functions
\citep{houlahan1,scf,vca} and Principal Component Analysis
\citep{heyer-turb}.  Statistical analyses produce many
interesting comparisons between and among data, but the physical
interpretation of the statistics can be complicated.  The most useful
applications of the statistical approach tend to be in comparative
measurements between two observational data sets or between
observations and a simulation \citep[e.g.][]{padoan-perseus}.

The segmentation and identification techniques are favored in the case
where the emission is thought to be comprised of physically important
substructures.  In molecular line astronomy, the classic examples of
the segmentation approach is the generation of GMC catalogs for the
inner galaxy where GMCs are identified as connected regions of
emission above a threshold intensity \citep{srby87,syscw}.
Unfortunately, the results of this approach is controlled by the
sensitivity and resolution of the data set\footnote{Sensitivity and
  resolution effects also contaminate analysis using statistical
  methods, but it is possible to correct for these effects
  \citep[e.g.][]{deltavar}.}.  The two applications of the segmentation
approach that have most shaped molecular line astronomy, particularly
with regards to the field of star formation, are the clump
identification algorithms of \citet{clumpfind} and
\citet{gaussclumps}.  The clumpy substructure of molecular clouds was
first identified by eye \citep{bs86} and this structure is thought to
be important at establishing the sites of star formation.
\citet{clumpfind} applied a watershed segmentation algorithm to
molecular cloud data to identify ``clumps'' within the cloud (the
now-famous CLUMPFIND algorithm).  The CLUMPFIND algorithm has spawned
many subsequent applications and its utility is discussed elsewhere
\citep{pineda-clumpfind}.  Where CLUMPFIND is driven by the structure
in the data and precludes finding overlapping objects, the GAUSSCLUMP
algorithm of \citet{gaussclumps} \citep[later revisited
  by][]{gaussclumps2} iteratively fits three-dimensional Gaussians to
data cube to identify structures in the data.  Both algorithms have
been used to define the mass spectrum of clumps within molecular
clouds, usually finding $\alpha \sim -1.5$ to$ -1.9 $ for $dN/dM
\propto N^{\alpha}$.  It should be noted that CLUMPFIND and GAUSSCLUMP
are not intended to produce the same partitioning of a data cube; they
adopt substantially different starting assumptions with a
corresponding difference in the results.  The results CLUMPFIND and,
to a lesser extent, GAUSSCLUMP are influenced by their user-defined
parameters and algorithmic design which are designed to mimic the
``by eye'' identification.

\section{The Dendrogram Technique}
\label{dendrograms}
The dendrogram technique presented here combines the robustness of the
statistical approach with the direct link to structure in the data
explored in the segmentation and identification approach.  The
analysis of dendrograms presented in HS92 highlights the utility of
the method at characterizing two-dimensional extinction maps using a
few simple statistics.

\begin{figure*}
\plotone{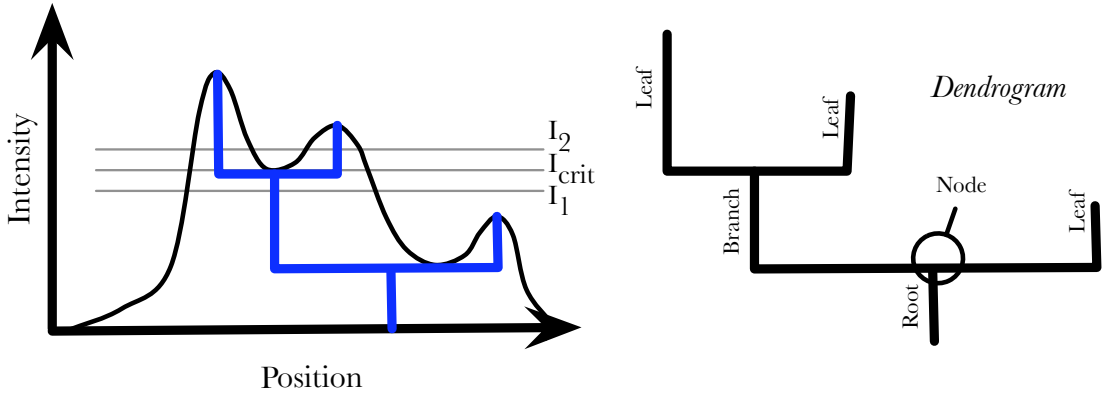}
\caption{\label{schematic} Schematic diagram of the dendrogram
  process.  The left panel shows a one-dimensional emission profile
  with three distinct local maxima.  The dendrogram of the region is
  shown in blue and repeated in the right panel where the components
  of the dendrogram are labeled.  The left-hand panel indicates three
  characteristic contour levels through the data.  Thresholding at
  $I_1$ produces a single object whereas thresholding at $I_2$
  produces two.  The level separating these two regimes is indicated
  as $I_{crit}$.}
\end{figure*}

We begin by considering images in general, without a specific
astronomical data type in mind.  This section emphasizes ideal data
where the presence of noise does not interfere with structure
identification.  As discussed in HS92, a dendrogram is a graphical
representation of the primitive structure within an image of arbitrary
dimension.  It is the skeleton of the object containing only
information about the structures and substructures within contour
diagram of the object.  A schematic of the dendrogram technique is
shown in Figure \ref{schematic} for a one-dimensional emission
profile.  If the emission profile were
thresholded\footnote{Thresholding is the mapping of a real-valued
  image to a binary image with all data above the threshold set to 1
  and all data below set to 0.} at level $I_{1}$ a single connected
region results.  In contrast, if the profile is thresholded at $I_{2}$
two distinct objects will result, corresponding to each of the local
maxima.  The level $I_{crit}$ represents the critical boundary between
these two regimes, below which two objects merge into a single
object. The dendrogram is a scheme to track the structure as a
function of contour level in the profile and thus it represents the
essential information about the structure of the object.  The
dendrogram also encodes where the composite object combines with the
third distinct object.

For two dimensional data, a common analogy is to think of the
dendrogram technique as a descriptor of a submerged mountain chain.
If the overlying water were drained away, first the peaks of mountains
would appear as isolated objects.  As successively more water is
drained, the peaks would merge together into larger objects.  The
dendrogram encodes information about which objects merged together and
at what contour levels they did so.  To plot a dendrogram of this data
we can flatten the two dimensional structure into one dimension but
doing so eliminates any positional information in the tree.  

A useful formalism for interpreting dendrograms in three dimensions is
to consider each point in the dendrogram as representing an isosurface
(3D contour) in the data cube at a given level.  If an arbitrary data
set is thresholded at a fixed contour level, it breaks up into one or
more distinct regions.  The bounding surfaces of these volumes are the
isosurfaces represented in the dendrograms, with each distinct surface
corresponding to a point in the dendrogram.  We identify the distinct
surfaces by the set of local maxima that they contain.  Over a range
of contour intervals with no mergers, thresholding the data at
slightly higher or lower level will produce the same essential
features, namely the same number of distinct regions containing the
same local maxima.  Hence, the dendrogram will be comprised of
vertical branches.  The length of these branches corresponds to the
range of contour levels over which a set of isosurfaces is unchanged
(though the actual volume will change).  There are specific contour
levels in the data above which a pair of volumes will be distinct and
below which the two volumes are joined.  We refer to these critical
levels as the {\it merge} levels.  Below the merge level, a single
isosurface contains both sets of local maxima that defined the
distinct surfaces above the merge level.  To represent this change in
the topology of the isosurfaces, we connect the two branches of the
dendrogram at the merge level.

\begin{figure}
\plotone{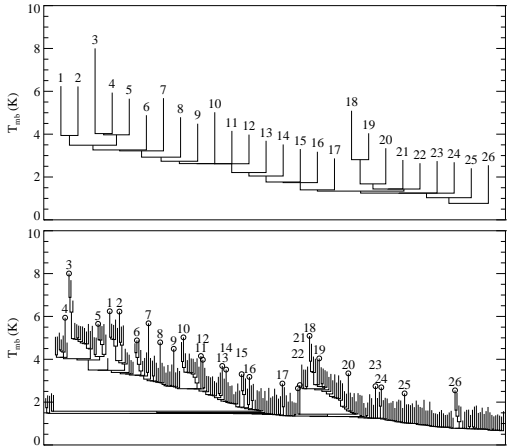}
\caption{\label{dendroex} Dendrograms of $^{13}$CO emission in L1448.
  The top panel shows the dendrogram of L1448 using the standard
  algorithm parameters.  The bottom panel shows the dendrogram
  after relaxing the conditions for noise suppression resulting in
  more independent leaves in the dendrogram (see the end of
  \S\ref{pruning}).  However, the basic structure of the dendrogram
  remains the same; the isosurfaces used in the top plot are a subset
  of those used in the bottom.  Each leaf of the dendrogram is labeled
  in the top plot and the corresponding leaf is identified in the
  bottom figure.  Leaves appearing in both dendrogram also have a
  circle at their tip in the bottom plot.}
\end{figure}

A sample dendrogram is shown in Figure \ref{dendroex} (top)
representing the $^{13}$CO ($1\to 0$) emission from the L1448 dark
cloud in Perseus \citep{complete-data}.  There is no spatial
information encoded in the $x$-axis of the plot but rather the
ordering of the leaves is chosen so that the branches of the
dendrogram do not cross.  This choice facilitates visualizing the
hierarchical structure in the data at the expense of retaining the
geometrical relations between the leaves.  The information on the
spatial relationships between the objects is retained in this analysis
and can be used to label maps with the regions of dendrograms they
correspond to, though it is not shown in the dendrogram.  The
construction of dendrograms including a discussion of the effects of
noise is discussed in more detail below.

\subsection{Determining the Leaves of a Dendrogram}
\label{pruning}

In the following sections, we specifically consider radio-line data
cubes in position-position-velocity space (PPV) with intensities given
in brightness temperatures (Kelvin).  Such observational data are
invariably contaminated by noise which interferes with the dendrogram
process.  The structure of the dendrogram is determined entirely by
the local maxima in the data.  A local maximum, by definition, has a
small region around it containing no data values larger than the local
maximum and, hence, a distinct isosurface containing only that local
maximum can be drawn.  The local maxima determine the top level of the
dendrogram, which we refer to as the {\it leaves}, defined as the set
of isosurfaces that contain a single local maximum.

In noiseless data, every local maximum in the data would correspond to
an actual emission feature in the data.  Unfortunately, in real data,
noise will mask the low-amplitude variations in the emission structure
resulting in spurious local maxima that do not correspond to real
structure in the data.  In the dendrogram method, we suppress the
effects of these noise fluctuations by rejecting local maxima that are
likely caused by noise.  

We describe our algorithm here in more detail considering, without
loss of generality, that we are examining only a single cloud of
emission such that a low-lying contour will contain all the emission
of interest in a cloud.  The initial leaves of the dendrogram are
selected by identifying all local maxima and then rejecting maxima
that are likely to be caused by noise.  We generate a list of all
local maxima by identifying all pixels in the image that have data
values larger than all of their neighbors over a box $D_{max}\times
D_{max}\times \Delta V_{max}$ in PPV space where $D_{max}$ and $\Delta
V_{max}$ are free parameters.  A non-trivial box size ($D_{max}$ and
$\Delta V$ greater than one pixel/channel) reduces the numbers of
candidates that must be checked for significance against our
noise-suppression criteria.  The algorithm becomes insensitive to
structure in the data cube on scales less than a box size.  If the box
is too large, significant structures are suppressed.  Since the
rejection of local maxima only simplifies the dendrogram by
considering a subset of the structurally defining features, reducing
the size of the noise-suppression box can be used to check if an
essential feature has been eliminated.

After the initial generation of local maxima, the set is then
decimated by removing local maxima that are likely to result from
noise.  For each pair of candidate maxima, we find the highest shared
isosurface that contains both maxima.  This isosurface is the merge
level, a high-dimensional analog of the contour level at the saddle
point shown in Figure \ref{schematic}.  For the merge level, we
calculate (a) the volume uniquely associated with each maximum and (b)
the difference in antenna temperature between the merge level and each
local maximum.  We remove any local maximum for which the volume of
the isosurfaces that contain only that maximum is less than some
minimum number of pixels ($N_{min}$, usually taken to be 4).
Furthermore, we only recognize a significant bifurcation in structure
when both local maxima are more than a given interval $\Delta T_{max}$
above the highest contour level that contains {\it both} of the
maxima, i.e. the level at which the objects merge.  Such a criterion
has been used previously in data cube analysis
\citep{outercat,m64-gmcs}: noise fluctuations will typically only
produce maxima with characteristic height $\sigma_{rms}$ so variations
significantly larger than this are nominally real.  If this criterion
is not fulfilled, we reject the lower of the two local maxima and
consider the emission profile to represent only a single object. We
note that the resulting dendrogram using a decimated set of local
maxima represents a set of isosurfaces that are a subset of the
isosurfaces that would be considered including all local maxima (see
Figure \ref{dendroex}). 

\begin{figure}
\plotone{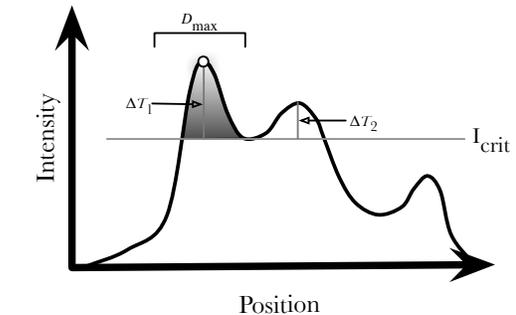}
\caption{\label{schematic2} Schematic diagram of the parameters that
  determine the decimation of local maxima.  The same profile as in
  Figure \ref{schematic} is used.  The local maximum indicated with
  the white point would be considered a valid local maximum if (a) it
  is the highest point in a window $D_{max}$ on either side of it (and
  an analogous width $\Delta V_{max}$ in velocity space), (b) the
  interval between the maximum and the highest merger level with a
  valid local maximum $\Delta T_{1}>\Delta T_{max}$ and if the number
  of pixels associated with the shaded region is larger than
  $N_{min}$.  These criteria restrict the analysis to the subset of
  local maxima that are most distinct. }
\end{figure}

Hence, the initial leaves of the dendrogram are determined by {\it
  four} free parameters: $D_{max}$, $\Delta V_{max}$, $\Delta T_{max}$
and the $N_{min}$.  By default, these are set to be $D_{max}=3$ and
$\Delta V_{max} = 7$~resolution elements, $\Delta
T_{max}=4\sigma_{rms}$ and $N_{min}=4$ pixels for independent pixels.
The bottom panel of Figure \ref{dendroex} shows the resulting
dendrogram for $D_{max}=1$ and $\Delta V_{max} = 3$~resolution
elements and $\Delta T_{max} = 0$.  Figure \ref{schematic2} is a
schematic diagram illustrating the definition of these parameters.
Changing $\Delta T_{max}$ results in the largest changes in the
dendrograms for typical radio line data since a larger fraction of the
local maxima fail the check against the contrast than any other noise
suppression criterion.  The default values represent a compromise
between sensitivity to dendrogram structure and algorithm performance.

Noise has an additional affect on the dendrogram, namely intensity
fluctuations can alter the levels at which two isosurfaces merge.  A
positive fluctuation can join two surfaces at a higher level than the
surfaces would join in the absence of noise.  We have modeled the
influence of the noise by comparing the merge levels of surfaces in a
model cube in the absence of noise to those with noise added.  We
find, in general, that the merge levels are uncertain on a scale of
$\sim 2\sigma_{rms}$ with some variation based on algorithm parameters
and the precise model used.  In addition, there is a bias towards
merging $\sim 1 \sigma_{rms}$ higher than the surfaces would merge in
the absence of noise.  The structure of the tree can only be
considered accurate for amplitude changes larger than $\sim
2\sigma_{rms}$, and for scales smaller than this, the branching order
may be transposed.  HS92 discuss this effect in some detail for 2D
images and resort to a coarse binning of their trees to measure tree
statistics (number of branches per node, etc.).  We do not present
tree statistics in this paper (see HS92), and we make an effort to
account for the influence of noise in our results.

\subsection{Constructing the Dendrogram}
\label{building}
Practically, the dendrogram of an $N$-dimensional intensity image is
constructed by first identifying the local maxima that will comprise
the top level of the dendrogram hierarchy (\S\ref{pruning}).  Then,
the data are contoured with a large number of levels.  For each
contour value beginning with the maximum level, the dendrogram
algorithm checks whether each pair of previously distinct regions have
merged together.  If so, the contour level and which surfaces merged
are recorded and the next contour level is considered.  We enforce
binary mergers: if three or more distinct objects merge into a single
object between one contour level and the next, we refine the
separation between contour levels so each merger involves only two
objects.  The dendrogram (tree diagram) is constructed by drawing
vertical segments corresponding to contour levels where the topology
of the surfaces are unchanged and connecting corresponding branches at
the levels where isosurfaces merge.

Both the identification of local maxima and the levels at which two
surfaces can be considered to be merged are influenced by the choice
of {\it connectivity} in the data set.  Practically, astronomical data
is pixellated into square (cubic) pixels.  The connectivity of the
data set is determined by the number of neighbors a given pixel is
defined to have.  In two dimensions, a pixel can have either four
neighbors (those pixels that share edges of the pixel) or eight
neighbors (those pixels that share corners or edges).  In three
dimensions, a cubic pixel can have either six neighbors (those pixels
which share a face with the cubic pixel) or 26 neighbors (those pixels
which share a face, edge or corner with a given pixel).  An
alternative definition in the three dimensional case considers a cubic
pixel to have 18 neighbors corresponding to those pixels which share a
face or an edge, but we emphasize 6- and 26-connectivity in the three
dimensional case to be analogous to the two-dimensional case
\citep[see][for further discussion]{clumpfind}.  Two points are in the
same region if a path can be drawn from one point to the other through
connected pixels which are all in the same region.  For our analysis,
we choose the minimum connectivity (4 neighbors in the 2D case, 6 in
the 3D), but we have experimented with the maximum connectivity.
Practically, the dendrogram changes by a small degree with
corresponding mergers, on average, occurring at higher contour levels
since it is ``easier'' for two regions to connect.

\section{Measuring Cloud Properties in Dendrograms}
\label{cloudprops}

Having developed a formalism where each point in the dendrogram
corresponds to a unique isosurface in the data, we calculate the
physical properties of the emission bounded by that isosurface.  We
can then use those physical properties to identify the relevant
features in the data cube.  Along branches of the dendrogram, the
properties tend to be continuous functions of the contour level, while
where two branches merge, the properties will change suddenly as a
result of the merged object containing more emission.  However, owing
to the difficulty in relating volumes in observed data space to
volumes in physical space, the measurement of properties from regions
of emission within the data cube is difficult to interpret (see
\S\ref{interp}).  In this section, we describe our methods for
estimating the size, line width, luminosity and mass contained within
each isosurface describe in the dendrogram as well as the
complications that arise in doing so.  

We calculate the macroscopic properties of the regions of emission
based on the moments of the volume weighted by the intensity of
emission coming from every pixel following \citet{props}.  The data
cube consists of a number of pixels that have sizes of $\delta x$,
$\delta y$, and $\delta v$ in the two spatial dimensions and the
velocity dimension, respectively. The $i$th pixel in the data cube has
positions $x_i$ and $y_i$, velocity $v_i$, and brightness temperature
$T_i$. We assume that the region under consideration is contiguous and
bordered by an isosurface in brightness temperature of value
$T_{edge}$, so that all of the pixels in the region have $T >
T_{edge}$ and the pixels outside the region have $T < T_{edge}$ or are
separated from the region by emission with $T < T_{edge}$.

We begin by rotating the spatial axes so that the $x$ and $y$ axes
align with the major and minor axis of the region, respectively. We
determine the orientation of the major axis using principal component
analysis.  The size of the region is computed as the geometric mean of
the second spatial moments along the major and minor axis. This is
$\sigma_{r}$, the root-mean-squared (RMS) spatial size:

\begin{equation}
\sigma_{r} (T_{edge}) = \sqrt{\sigma_{maj} (T_{edge})~\sigma_{min} (T_{edge})}
\end{equation}

\noindent where $\sigma_{maj} (T_{edge})$ and $\sigma_{min}
(T_{edge})$ are the RMS sizes derived from the intensity-weighted
second moments along the two spatial dimensions.
\begin{equation}
\label{moment}
\sigma^2_{maj} = \frac{\sum_i w_i \left(x_i-\langle x\rangle\right)^2}
{\sum_i w_i},
\end{equation}
where we have assumed the major axis lies along the $x$ coordinate and
the sum runs over all pixels within the isosurface ($T>T_{edge}$).
The weights in the moment are usually set to the brightness
temperature of each pixel: $w_i=T_i$.  This particular functional form
for the cloud size is used since it has been used in previous
observational studies \citep{srby87} and explored in depth by
\citet{bertoldi-mckee} with respect to inclination, aspect ratio, and
virialization.  We define a factor $\eta$ that relates the
one-dimensional RMS size, $\sigma_r$, to the radius of a spherical
cloud $R$: $R=\eta \sigma_r$.  We take $\eta=1.91$ for consistency
with \citet{srby87} and \citet{props}; the value of 1.91 merely
reflects the correction of the moment to the radius for the typical
concentration of emission found in molecular clouds.  The velocity
dispersion ($\sigma_v$) is calculated as the second moment of the
velocity axis weighted by the data values, analogous to the size
measurement.  The flux of the region is the sum (zeroth moment) of all
the emission in the region: $F=\sum_i~T_i~\delta \theta_x~\delta
\theta_y~\delta v$.  To convert the flux to a luminosity, we must
assume a distance to the region.  For a cloud at a distance of $d$ (in
parsecs), the physical radius will be:
$R_{\mathrm{pc}}=R_{\mathrm{rad}}d$ and the luminosity will be $L = F
d^2$ where the flux is measured in units of K~km~s$^{-1}$~sr.  For CO
data, we calculate the mass of the region, we scale by a linear
CO-to-H$_2$ conversion factor (for intensities on the main beam
temperature scale):
\begin{equation}
\frac{M_{\mathrm{Lum}}}{M_{\odot}} = \frac{X_{\mathrm{CO}}}{2 \times
  10^{20} [\mathrm{cm}^{-2}/(\mathrm{K~km~s}^{-1})]} \times
4.4~\frac{L_{\mathrm{CO}}}{\mbox{K km s}^{-1}\mbox{ pc}^2} \equiv
4.4~X_2~L_{\mathrm{CO}},
\end{equation}
\noindent where $X_{\mathrm{CO}}$ is the assumed CO-to-H$_2$
conversion factor. This calculation includes a factor of 1.36 (by
mass) to account for the presence of helium. Including helium is
necessary to facilitate comparison with the virial mass, which should
reflect all of the gravitating mass in the cloud.  We have adopted a
fiducial value of the CO-to-H$_2$ conversion factor of
$X_{\mathrm{CO}}=2\times 10^{20}\mbox{ cm}^{-2} (\mbox{K km
  s}^{-1})^{-1}$ based on $^{12}$CO($1 \to 0$) observations in the
Milky Way \citep{sm96,dht01} and express changes relative to this
value in terms of the parameter $X_2$.

For each property, we can estimate the uncertainty in the property
caused by the noise in the data set.  Assuming the coordinate axes are
well-defined, the uncertainties are algebraically propagated through
the formulas for the physical properties.

\subsection{Physical Interpretation}
\label{interp}
\begin{figure*}[ht]
\plotone{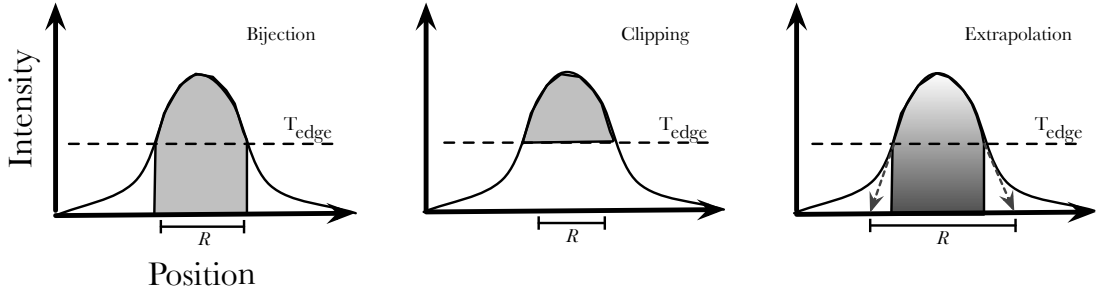}
\caption{\label{schematic-3par} Graphical summary of the three
  paradigms for interpreting isosurfaces of emission investigated in
  this work.  The figure shows the same one-dimensional emission
  profile and a contour level for each of the three cases.  The shaded
  area shows the emission used to compute cloud properties (and is
  proportional to the luminosity).  A bar is shown below the position
  axis indicating the relative extent or a moment-based size
  measurement under the three schemes.  The standard interpretation of
  isosurfaces is the bijection scheme where elements in observational
  space correspond directly to objects in physical space.  In the
  clipping paradigm, only emission above the contour level is
  associated with an object.  In the extrapolation scheme, all the
  elements above the contour level are used to infer the behavior of
  the calculated properties in an extrapolation to the zero intensity
  isosurface. A similar set of characterizations would hold if
  velocity were the coordinate axis and the line width was measured. }
\end{figure*}

The major difficulty in using the above calculated properties directly
is that it is difficult to ascribe meaning to a region of emission
defined by an isosurface.  There is substantial concern that the naive
association of a closed object in PPV space with an object in physical
space, particularly as defined by contours of intensity in a data
cube, may be inaccurate \citep{osg01}.  From the observer's
perspective there seem to be three possible interpretations for the
emission in the data cube.  Each of these interpretations leads to
different sets of cloud properties and yields different results when
applied to the same data set.  These interpretations all revolve
around determining what the appropriate values of the antenna
temperature weights used in moments of the emission ($w_i$) should be
(e.g., those in Equation \ref{moment}).  We graphically summarize the
three ``paradigms'' for measuring the properties of isosurfaces in
Figure \ref{schematic-3par}.

\subsubsection{The Bijection Paradigm}

The calculations of properties sets the weights to the native values
of brightness temperature drawn from the data cube $w_i=T_i$.  This
assumption essentially maps PPV space to physical space (i.e. three
spatial dimensions) in a 1-to-1 fashion (one pixel in the data cube
corresponds to single volume in the cloud).  To the extent that this
is true, this is the correct thing to do.  This result is the closest
parallel to the CLUMPFIND algorithm which associates clumps of
emission with clumps of density in physical space.  Under the
assumption of uniform excitation conditions, an isosurface of
brightness corresponds to a surface of constant opacity and hence of
constant column density.  In the physical regime where higher column
densities are associated with higher physical densities, the bijection
paradigm may be ideally suited for measuring cloud properties.

A bijection may be inappropriate because of two effects: the first is
the superposition of multiple, distinct objects along the line of
sight that have the same velocity.  The second is that a given volume
likely contributes emission at multiple velocities due to an
intrinsically broad line profile.  Both of these effects cause the
bijection to be flawed: the first means that a given pixel contains
emission from multiple objects and the second means that any given
volume appears in multiple pixels.  We can attempt to correct for
either of these effects, but not both. 

\subsubsection{The Clipping Paradigm}

\begin{figure*}
\plotone{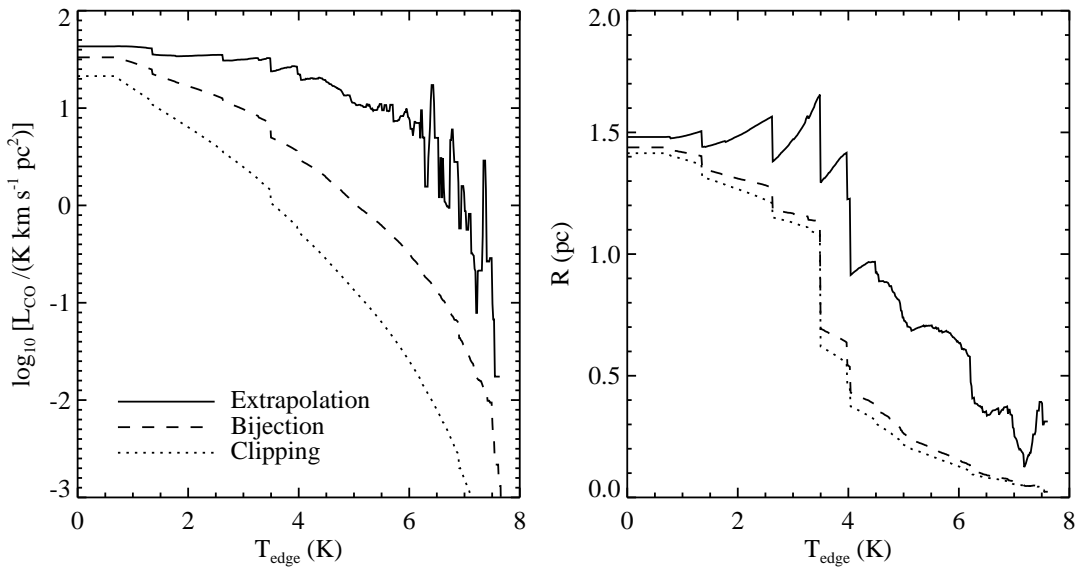}
\caption{\label{propplot} The behavior of calculated properties as a
  function of threshold level in the dendrogram for the set of
  isosurfaces containing the maximum of the L1448 data cube.  The
  behavior of the cloud properties along the highlighted path are
  shown for the luminosity (left) and radius (right).  Three curves
  are shown in each of these panels representing three possible ways
  of calculating the cloud properties at a given $T_{edge}$.  The
  bijection paradigm is shown as a dashed curve ($w_i=T_{i}$).  The
  dotted curve depicts the clipping treatment ($w_i=T_{i}-T_{edge}$),
  and the solid curve shows the result of extrapolating the emission
  to the zero intensity isosurface.}
\end{figure*}

In this approach, the region is considered to represent a discrete
object superimposed on a background of brightness $T_{edge}$.  This
approach assumes that any emission that can be associated with other
objects, by drawing a lower contour, is not associated with the object
at all.  In this case, the properties of the clouds should be
calculated using weights $w_i=T_i - T_{edge}$.  The resulting values
of some representative properties are shown in Figure \ref{propplot}
for the set of isosurfaces containing the maximum in the L1448 data
cube.  In general, the clipping tends to reduce all the properties,
but affects the luminosity most significantly.  This assumption is
very conservative and the correct value would be derived using a
weight value intermediate between 0 and $T_{edge}$.

\subsubsection{The Extrapolation Paradigm}

In this paradigm, the properties of the region are extrapolated to the
zero-intensity isosurface.  The extrapolation corrects for the fact
that some of the emission arising from the object is not contained
within the contour drawn in PPV space.  Instead of quoting the
properties of the measured region, the extrapolation reports {\it the
  properties implied for the entire region as inferred from the part
  found above $T_{edge}$}.  An analogy for this correction is that we
are predicting the underwater shape of an island volcano from the
visible region above the water.

The extrapolation is carried out by considering the behavior of a
property, say $R$, as a function of $T_{edge}$.  For a given
$T_{edge}$, we extrapolate $R(T_{edge})$ to a value of
$T_{edge}=0~\mbox{K}$ based on the behavior of $R(T'_{edge})$ for all
$T'_{edge} > T_{edge}$.  This method is described in more detail in
\citet{props}, in particular in their Figure 2.  In short, the second
moments are linear extrapolations for data above $T_{edge}$ whereas
the zeroth moments are quadratic extrapolations.  The behavior of the
extrapolation can be traced on Figure \ref{propplot}.  The value of
the extrapolated radius is always larger than for the other two
paradigms and the margin of difference is most substantial at large
contour values since the range of the extrapolation is the largest.
At small values of $T_{edge}$, the extrapolated value oscillates around
$\sim 1.5$ pc, the final radius of the cloud.    

It is the latter quadratic extrapolation that produces the noise on
the black curve in the left panel of Figure \ref{propplot}.  While
this method corrects for the emission associated with the object at
low intensity values, it effectively adds emission to the objects so
that the sum of the extrapolated objects from high intensity values
may be larger than the amount of emission contained in the data cube.
Hence, ratios such as the virial parameter (\S\ref{virparam}) will be
more accurate in this assumption than will integrated properties like
radius or mass as a function of contour level.

For the extrapolation paradigm, the dominant source of errors can be
the actual data used in the fit and the errors can be assessed by
bootstrapping the data used in the extrapolation \citep{numrec}.  See
\citet{props} for more details.

\subsection{The Virial Parameter}
\label{virparam}
We adopt the virial parameter as defined in \citet{mckee-vt} as a
diagnostic of the energetic state of the regions in the dendrogram.
In this case, the virial parameter $\alpha$ is defined as:
\begin{equation}
\alpha = \frac{5 \sigma_v^2 R}{M_{\mathrm{Lum}} G} =  
\frac{5 \eta \sigma_v^2 \sigma_r}{4.4 X_2 L_{\mathrm{CO}} G}
~\mbox{,}
\end{equation}
where $L_{\mathrm{CO}}$ is measured in units of
K~km~s$^{-1}$~pc$^{2}$.  For $\alpha < 2$, the object is
self-gravitating {\it in the absence of other forces.}  Magnetic
fields, surface pressures and bulk motions will all affect the
dynamical state of the cloud.  Since such terms are not readily
measurable, we must adopt this simple estimate for the dynamical state
with the understanding that it is only an approximation.  The utility
of the diagnostic is most likely in a relative sense rather than an
absolute one.  We should regard regions with $\alpha \lesssim 2$ as
regions where significant amounts of gravitational potential (mass)
are found with comparably little kinetic energy so that gravity is
likely important.  In the remainder of the paper, we refer to such
regions as ``self-gravitating'' though the description is subject to
the caveats above.

One concern that arises is the meaning of the virial parameter under
the three different approaches to calculating the region properties
that go into the virial parameter.  When adopting the bijection
approach, measuring the virial parameter for emission contained above
a given contour could be inaccurate since the omission of the
``wings'' of the cloud would affect the size and line width
measurements more than the luminosity measurement.  Under the clipping
approach, the assumption that none of the emission below a given
contour level is associated with the object results in similar size
and line widths while dramatically reducing the luminosity.  As a
result, this assumption likely overestimates the virial parameter for
the region.  Finally, the virial parameter measured in the
extrapolated case characterizes the dynamical state of the region
implied by the emission found above a given contour level.  The
extrapolation method is most useful for characterizing objects for
which the zero intensity isosurface is a meaningful boundary
(i.e. discrete clouds rather than substructure within clouds).  Given
these considerations and our emphasis on using virial parameter to
estimate the dynamical state of structures in the data we adopt the
bijection scheme for characterizing substructure, basically
interpreting the isosurfaces in the data as corresponding to nested
regions of successively higher (column) density.  When identifying
clouds or other objects for which the zero intensity isosurface is a
more meaningful boundary, we adopt the extrapolation paradigm.  We
discuss the influence of these choices further in
\S\ref{l1448-threepar} in the application to observed data.

\section{The Hierarchical Substructure of a Molecular Cloud}
\label{l1448}
In this section, we apply the dendrogram method to two data sets
and demonstrate useful statistics for the characterization of the trees
and the tree-based properties.  

\subsection{L1448 COMPLETE Data}
\begin{figure*}
\plotone{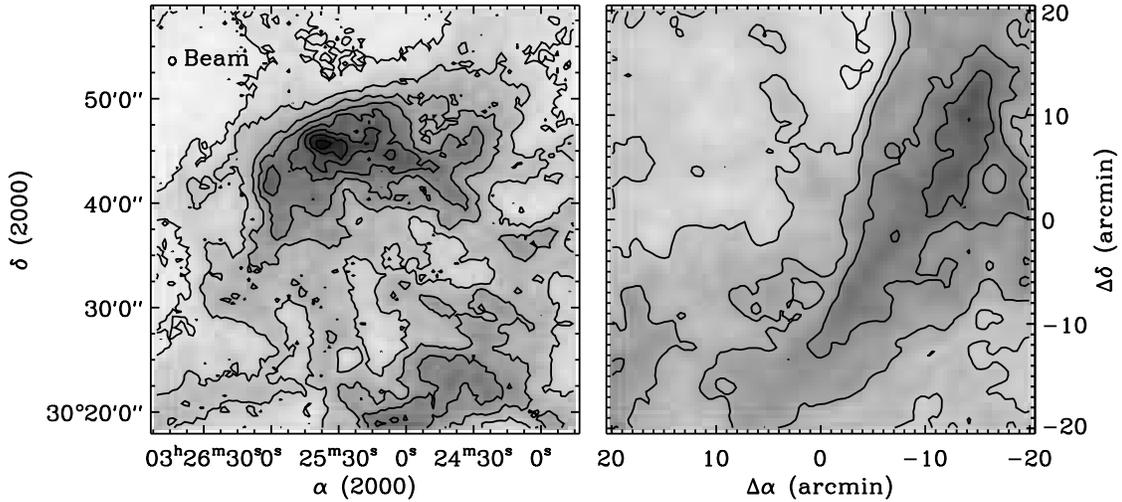}
\caption{\label{momentmap} Integrated intensity maps of $^{13}$CO
  emission from L1448 (left) and the simulation of
  \citet{padoan-perseus} (right).  In both images, the gray scale runs
  linearly from 0 to 18 K km~s$^{-1}$ on the $T_{\mathrm{mb}}$ scale
  and the contours run from 2-16 K km~s$^{-1}$ in intervals of 2 K
  km~s$^{-1}$ in the left-hand panel and from 4-10~K~km~s$^{-1}$ in the
  right-hand panel.  The two maps come from data cubes with the same
  native resolution and noise levels.  
}
\end{figure*}

\begin{figure*}
\plotone{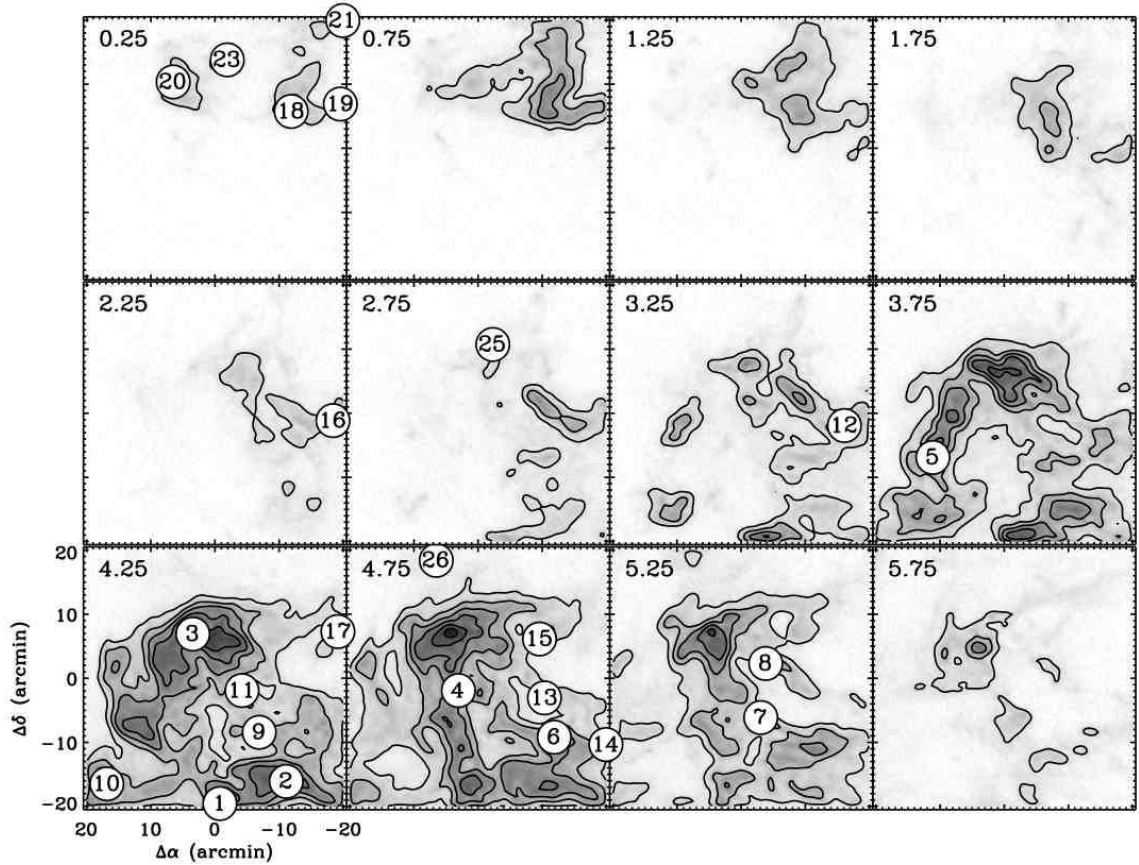}
\caption{\label{chanmaps} Selected channel maps of the L1448 region.
  Contours follow the grayscale image with a contour interval of 1~K
  on the $T_{\mathrm{mb}}$ scale beginning at 1~K.  The LSR velocity
  of each map is indicated in the upper left-hand corner of each
  panel.  The channel maps reveal the low velocity feature
  ($v_{\mathrm{LSR}}\sim 0.5\mbox{ km s}^{-1}$) not otherwise
  discernible in the integrated intensity map.  The 3D positions of
  the leaves of the dendrogram are indicated with the leaf number
  shown in Figure \ref{dendroex}, shown in the closest channel map to
  their actual position.  }
\end{figure*}

Our primary observational data set for demonstrating the dendrogram
method is a section of the Coordinated Molecular Probe Line Extinction
Thermal Emission\footnote{COMPLETE;
  \url{http://www.cfa.harvard.edu/COMPLETE/}} survey's $^{13}$CO map
of Perseus \citep{complete-data}, centered on the L1448 star-forming
region.  The data cube spans a square region 40$'$ on a side which
projects to a region 3.1~pc~$\times$~3.1~pc at the distance of Perseus
\citep[260 pc,][]{cernis93}.  The data have an angular resolution of
46$''$ and are sampled with 23$''$ pixels.  Since the observational
methods produce non-uniform noise across the map, we add appropriately
correlated noise to the original data to produce a map with
spatially-uniform noise rms of ($\sigma_{rms}=0.3$~K on the main beam
scale).  The data cube has a velocity resolution of
$0.066$~km~s$^{-1}$, sampled every 0.066~km~s$^{-1}$, and spans 40
km~s$^{-1}$, but the emission from L1448 only spans a 10 km~s$^{-1}$
section of the data.  An integrated intensity map of the cloud is
shown in Figure \ref{momentmap} and channel maps are presented in
Figure \ref{chanmaps}.  The channel maps highlight the presence of a
low velocity feature not otherwise discernible in the integrated
intensity maps ($v_{\mathrm{LSR}}\sim 0.5\mbox{ km s}^{-1}$) .  The
main and low velocity features are contained within a single connected
isosurface for contour levels $< 1.5~\mathrm{K}$.  Individual clumps
corresponding to the branches of the dendrograms (see below) can be
seen as well as the rough positions of the local maxima used in our
analysis of the region (\S\ref{l1448dendro}).

\subsection{Turbulent Simulation}
For comparison to the data, we also analyze simulation data from
\citet{padoan-perseus}.  The data are taken from their simulated
$^{13}$CO emission maps generated from a 6 pc simulation box with a
mean density of $n=10^{3}\mbox{ cm}^{-3}$.  The simulation is
conducted using the {\it Enzo} code \citep{enzo} to simulate a
1024$^3$ box using MHD with an initially uniform density and periodic
boundary conditions.  The mean Mach number in the simulation is
$\mathcal{M}=6$; the simulation is isothermal and turbulence is driven
in Fourier space at large scales.  

For comparison with observed data, \citet{padoan-perseus} generated a
simulated $^{13}$CO data set using a Monte Carlo radiative transfer code
using the density and velocity distributions of the simulated material
in a snapshot (i.e.~radiation is not included in the time propagation
of the simulation).  We extracted a trial data set matching the
spatial extent of the observations from the full simulation box.  Our
selection included the section of the box that contained the most
compact identifiable feature of emission.  The trial data were
convolved to the resolution of the FCRAO maps from COMPLETE and
resampled in position and velocity to match the pixel size of the
observational data.  Spatially correlated noise, mimicking the noise
in the FCRAO map, was added to the simulation data to produce the same
underlying noise rms in both data cubes.  Both the simulation and the
observed data cubes will affected in the same manner by edges,
resolution and noise.  The only differences between the two data sets
should be found in the detailed structure of the emission.  The
simulation was compared favorably to the full COMPLETE observational
data set in \citet{padoan-perseus} based on similarities in the
turbulent power spectrum.  However, the authors of that study
emphasize that the simulation is not intended to simulate specific
conditions within a molecular cloud.  We focused our comparative
analysis with this simulation to illustrate the utility of dendrograms
even though the simulation box may not be an excellent simulacrum of
the L1448 region in particular.

\subsection{The $^{13}$CO-to-H$_2$ Conversion Factor}
\label{xfac}
\begin{figure}
\plotone{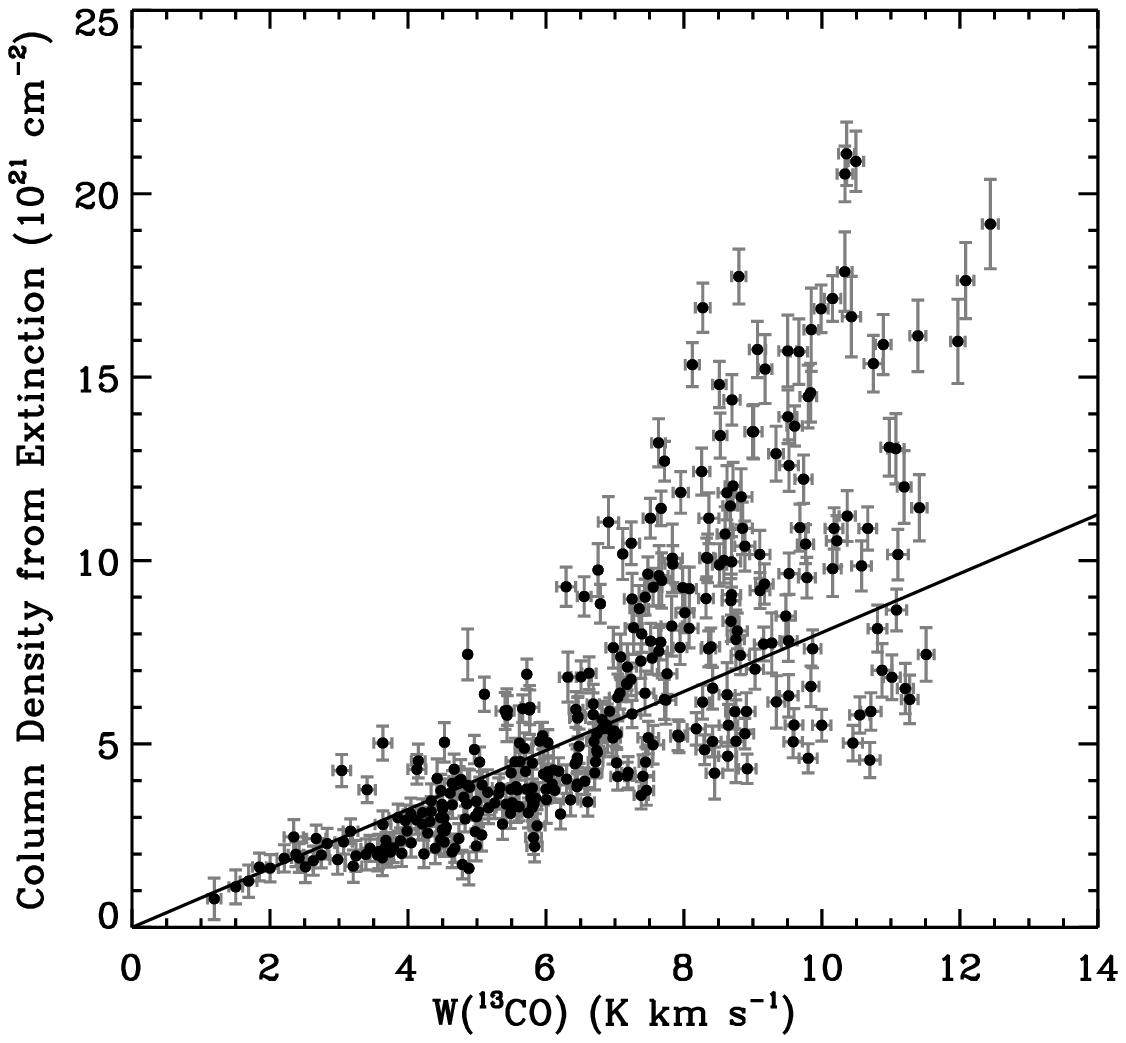}
\caption{\label{avco} Relationship between integrated intensity for
  $^{13}$CO and the H$_2$ column density implied by reddening in the
  near infrared.  A line representing the mean ratio between the two
  quantities is shown.}  
\end{figure}

We determine the scaling between $^{13}$CO luminosity and molecular
cloud mass by comparing the integrated intensity of the $^{13}$CO
emission to the extinction implied by the reddening of background
stars.  We use the extinction map for the L1448 region derived from
deep $JHK$ Calar Alto observations of the L1448 region using the Near
Infrared Color Excess Revisited \citep[NICER][]{nicer} technique
\citep{cloudshine}.  The $^{13}$CO integrated intensity map is
convolved and regridded to match the resolution ($48''$) and
astrometry of the extinction map.  The extinction map saturates above
$A_V\sim 22$ mag, and we ignore the 13 pixels with missing data in the
analysis.  Figure \ref{avco} shows the implied column density
\citep[assuming $N(\mathrm{H}_2)/A_{V}=9.4\times 10^{20}~\mbox{
    cm}^{-2}$ and $R_V=3.1$,][]{dust2gas} as a function of the
integrated intensity.  We calculate a $^{13}$CO-to-H$_2$ conversion
factor of
\begin{equation}
X_{13\mathrm{CO}} = 8.0 \times 10^{20},
\frac{\mathrm{cm}^{-2}}{\mbox{K km s}^{-1}}
\end{equation}
based on the mean of $N(\mathrm{H}_2)/W(^{13}\mathrm{CO})$ weighted by
the inverse variance of the column density estimates.  As seen in
Figure \ref{avco}, the single conversion factor is nowhere an
excellent approximation of the data, but represents an adequate
mapping between CO emission and column density over the entire range.
Since the conversion factor will be ultimately applied to individual
channels, only a simple ratio is appropriate for the relationship;
including complications such as non-linearities or a constant offset
would bring up ambiguities in translating from $^{13}$CO emission to
mass for individual channels in the data cube.  The simple ratio
systematically underestimates the column density for high brightness
regions where the $^{13}$CO line saturates.  As such, estimates for
the virial parameter are likely overestimates in these regions.  In
terms of the mass calculations presented in \S\ref{cloudprops}, $X_2 =
4.0$.  This is comparable to the results of more sophisticated
analyses of the conversion factor \citep[$X_2=2.1$;][]{pineda-abund}
though their complex analysis is only applicable for total line intensity.

\subsection{The Dynamical State of L1448}
\label{l1448dendro}
\begin{figure}[ht]
\plotone{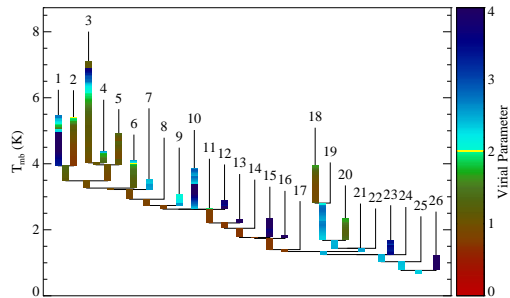}
\caption{\label{l1448color} Dendrogram of the L1448 region with
  branches of the dendrogram colored according to the virial parameter
  at each point.  Virial parameter data are suppressed where the
  errors are larger than 50\%.  Several of the leaves of the
  dendrogram show evidence for self-gravitation as do larger
  structures in data.  Since physical properties are calculated for
  the isosurfaces corresponding to the vertical branches of the
  figure, the horizontal branches of the dendrogram have no data
  reported.}
\end{figure}

We generate a dendrogram of the L1448 region with the $^{13}$CO data
using the methods discussed in \S\ref{dendrograms}.  We identify local
maxima over a box that is 2 beam widths on a side and 5 channels deep
in the data cube.  From this set, we eliminate redundant local maxima
that are less than 1.2~K ($4\sigma_{rms}$) above the level at which
the surfaces containing that local maximum merge with other
structures.  We remind the reader that this decimation preserves the
overall structure of the dendrogram and is considering a
representative subset of the topologically important surfaces in the
analysis (Figure \ref{dendroex}).  We calculate the virial parameter
for the region as discussed in \S\ref{virparam} using the bijection
scheme to calculate isosurface properties.  In Figure
\ref{l1448color}, we plot the dendrogram of the region, color coding
points on the dendrogram with the corresponding value of the virial
parameter.  We have calculated the errors in the virial parameter and
suppress reporting any values where the formal errors in the virial
parameter are larger than 50\%.  Figure \ref{l1448color} shows that
several leaves of the dendrogram show evidence for self-gravitation on
small scales associated with individual local maxima.  Note that the
left-hand branches (leaves 1-17) of the dendrogram appear
self-gravitating, but for contour levels $\lesssim 1.5\mbox{ K}$ where
the left and right branches merge, the ensemble properties of the
object revert to being unbound.  This change in dynamical state shows
the main complex of L1448 is dynamically distinct from the low
velocity feature at $v_{\mathrm{LSR}}\sim 0.5 \mbox{ km s}^{-1}$
indicated by the branch on the right (leaves 18-22).

In general, the tops of the dendrogram leaves do not appear as
self-gravitating objects in this analysis.  However, it is precisely
in these regions where the $^{13}$CO tracer saturates so that these
isosurfaces correspond to relatively more mass per unit brightness
than our simple conversion factor admits and hence will be more
tightly bound than we measure.  Owing to the difficulties in assessing
the dynamical state of these small objects, we emphasize the larger
scales for self-gravitation in our analysis.  These large
scales are where CLUMPFIND and GAUSSCLUMP fail to probe and
dendrograms provide novel insight.

\begin{figure}[h]
\plotone{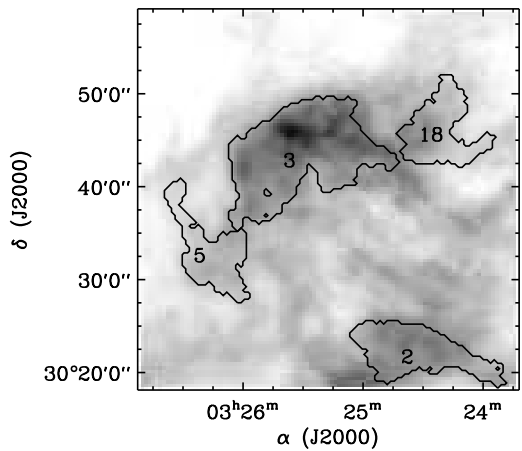}
\caption{\label{objmap} Integrated intensity image of the L1448 region
with the location and extent of the self-gravitating leaves in the
dendrogram indicated.  Leaf 3 contains most of the star-formation
currently occurring in the region.  Leaf 18 is the dynamically
distinct feature at low $v_{\mathrm{LSR}} (\sim 0.5$~km~s$^{-1}$).}
\end{figure}

\begin{figure*}[ht]
\plotone{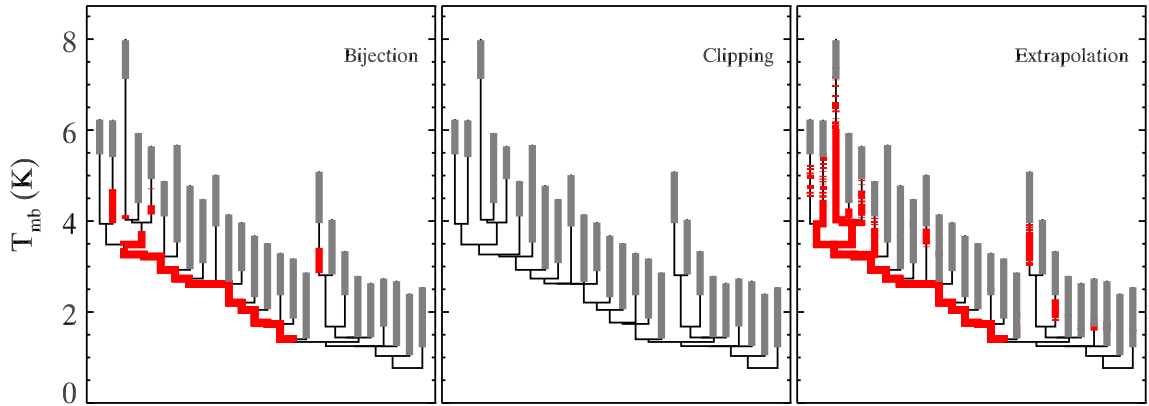}
\caption{\label{l1448vp} Self-gravitating objects in the L1448
  dendrograms based on the three different property calculation
  paradigms presented in \S\ref{interp}.  For each dendrogram, regions
  with $\alpha \le 2$ are shaded in red and regions where the data
  quality prohibit calculation of the properties are highlighted in
  gray.  The extent of the self-gravitating regime in parameter space
  depends on the paradigm adapted though many qualitative features are
  shared between the bijection and clipping paradigms.}
\end{figure*}

The multiscale analysis of the virial parameter allows us to define
objects that are potentially physically relevant to the star formation
process.  We identify objects based on the criterion that self-gravity
makes a significant contribution to their internal energetics.  If we
define a threshold for significant self-gravity, namely $\alpha \le
2$, we find ``interesting'' objects on a variety of scales.  Applying
this criterion to the virial parameter data shown in Figure
\ref{l1448color} results in the dendrogram shown in Figure
\ref{l1448vp} (left panel) where branches with $\alpha \le 2$ are
shaded.  Nearly all of the left-hand branch of the dendrogram
corresponds to a self-gravitating object indicating the importance of
self-gravity over the entire L1448 region. There are also three
distinct sub-branches inside L1448 that also show self-gravitation.
Figure \ref{objmap} shows the locations and spatial extent of the four
leaves that show evidence of self-gravitation in the data cube (2,3,5
\& 18).  The central, star-forming section of L1448 is contained in
leaf 3.  Also interesting are the several branches for which there are
large regions with reliable measures of the virial parameter which are
not self-gravitating.  Referring to Figure \ref{l1448color}, these
branches have $\alpha\gg 2$.  Because of the minimal influence of
self-gravity on these structures, we contend that these branches
correspond to transient or pressure-confined structures in the
physical data.

\subsection{The Dynamical State of the Turbulent Simulation}
\begin{figure}
\plotone{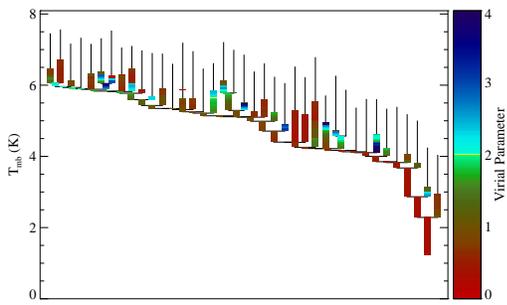}
\caption{\label{simdendro} Dendrogram of the simulation data cube
  colored according to the virial parameter at each point.  Virial
  parameter data are suppressed where the errors are larger than 50\%.
  Nearly all of the structure in the simulated data cube corresponds
  to self-gravitating objects with a few leaves of the dendrogram
  representing unbound objects.}
\end{figure}

We have repeated the dendrogram analysis for the turbulent simulation
using the same algorithm parameters to establish local maxima and
contour the data. We adopt a $^{13}$CO-to-H$_2$ conversion factor of
$X_2=10.9$ based on analysis of the simulated $^{13}$CO data with
respect to the simulated column density, using the same analysis as
was used in in \S\ref{xfac}.  The dendrogram presented in Figure
\ref{simdendro}, which can be compared to the observed data in Figure
\ref{l1448color}.  The simulated and observed data cubes have similar
numbers of leaves (local maxima) in their respective data volumes (39
in the simulation vs.~26 in the observations).  The span of antenna
temperatures are similar, though most of the mergers in the simulated
data cubes occur at higher levels than in the observed data.  The
principal difference between the two dendrograms is that far more of
the simulated data cube corresponds to self-gravitating objects than
do the actual observations.  Regardless of the applicability of the
dendrogram interpretations, the analysis illustrates a stark
difference in the data cubes.  The difference in dynamical states
arises from amount of mass in the two data cube.  Scaling the total
emission in each data cube by the respective conversion factors shows
there is $\sim 4$ times as much molecular mass in the simulation cube
as there is in the L1448 region, but this extra mass is spread over a
similar line width and spatial extent.  As a result, self-gravity
would play a stronger role in the simulated data cube.  The simulation
does not include the effects of gravity although our basic analysis
suggests that self-gravity would have a significant influence on the
simulated region.

\subsection{Interpretation of Dendrogram Properties}
\label{l1448-threepar}
The previous section discusses the physical meaning of the dendrograms
under the assumption that the ``bijection'' paradigm holds relating
objects in observational and physical spaces.  Previously
(\S\ref{interp}), we presented two other possibilities for relating
the observed and physical domains, namely the clipping and
extrapolation paradigms.  We repeat the calculation of the virial
parameter in L1448 for these two possibilities and present the results
alongside the bijection results in Figure \ref{l1448vp}.  The
extremely conservative clipping paradigm finds no self-gravitating
structure in the entirety of the L1448 cloud.  Given that simple
calculations suggest that the region has a virial parameter of $\alpha
\sim 2$ and the presence of star forming clumps at the smallest
scales, we conclude that the clipping paradigm is overly conservative
and the small structures have more mass than are accorded to them.
The extrapolation paradigm finds more self-gravitating structure in
the map than the bijection, which is expected since the extrapolation
corrects the luminosity by a larger factor than the radius and the
line width (see Figure \ref{propplot}).  However, it is interesting to
note that the same qualitative behavior is present in the extrapolated
results as are seen in the bijection analysis.  In particular, the
analysis finds two dynamically distinct regions in L1448 corresponding
to the left- and right-hand branches of the dendrogram.  However, the
extrapolation results assume that every object should have a
brightness profile that runs continuously from the peak value to the
zero brightness isosurface, and it may not be applicable in this case.
For simplicity, we utilize the bijection scheme for calculating the
properties of object substructure although extrapolation may be
appropriate in cases where there should be no background emission (see
\S\ref{gmcs}).

\begin{figure}
\plotone{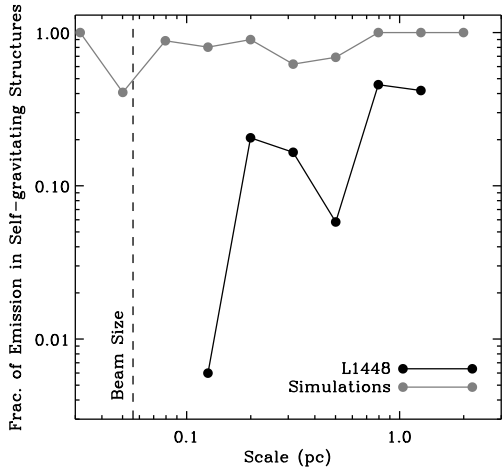}
\caption{\label{virscale} Fraction of emission contained within
  isosurfaces corresponding to self-gravitating objects as a function
  of size scale for the L1448 $^{13}$CO($1\to 0$) data and the
  simulated $^{13}$CO observations.  At small scales, very few objects
  are self-gravitating, but this fraction grows for larger size
  scales.  The simulations have roughly constant fractions of
  self-gravitation across size scales.}
\end{figure}

\subsection{The Scale of Self-Gravity}
Many previous authors have identified self-gravitating substructure in
their analysis of molecular emission.  What, then, makes the
dendrogram analysis novel?  Using by-eye identification
\citep{bs86,bertoldi-mckee} or automated algorithms such as CLUMPFIND
or GAUSSCLUMP invariably finds that the most massive objects on the
smallest physical scales sampled by the observations are closest to
being self-gravitating \citep[Figure 3 in][]{bertoldi-mckee}.
However, segmentation tends to identify structures on small scales (a
few resolution elements), ignoring the objects at larger scales which
comprise the superstructure of the molecular cloud.  Dendrogram
analysis avoids segmentation and naturally includes these larger
scales since lower valued isosurfaces encompass more emission with
larger spatial extents.

The multi-scale nature of the analysis naturally admits a study such
as that shown in the Figure \ref{virscale} which displays the fraction
of emission on given scales in L1448 and in the turbulent simulation that
has $\alpha < 2$.  We construct the diagram by measuring the virial
parameter as a function of size scale for all the isosurfaces in the
data cube.  This fraction is defined as the sum over all isosurfaces
\begin{equation}
f(R) \equiv \frac{\sum_i \{L_i | R_i \in [R,R+\Delta R], \alpha_i \le
  2\}}{\sum_i \{L_i | R_i \in [R,R+\Delta R]\}}
\end{equation}
where $L_i$, $R_i$ and $\alpha_i$ are the luminosity, radius, and
virial parameter of the $i$th isosurface. We bin the virial parameter
data into bins of $\Delta R=0.2$ dex in size scale and calculate the
fraction of luminosity in each bin contained within isosurfaces
corresponding to self-gravitating objects.  This calculation
illustrates that only a small fraction of the structure at small
scales in L1448 corresponds to self-gravitating objects and that
fraction grows at larger scales.  The saturation of the $^{13}$CO line
in small, bright regions makes this measurement a lower limit since
more of the leaves may correspond to self-gravitating objects than are
recovered in this analysis.  In contrast, self-gravity is important
for nearly all structure at all scales in the simulated observations.
Several factors may contribute to this discrepancy.  Incomplete
physics in the simulation (lack of self-gravity, too high of density)
or the incomplete synthesis of spectral line maps (no depletion
assumed) may give discrepant results.  Alternatively, the comparison
may be flawed and the section of the simulation box used may be not be
appropriate for comparison to L1448.

\subsection{The Size-Line Width Relationship in L1448}

In addition to identifying sets of isosurfaces that correspond to
self-gravitating objects, the dendrogram technique also provides
another way to probe the size-line width relationship on intermediate
scales inside the molecular clouds.  The Principal Component Analysis
methods developed in \citet{heyer-pca} and \citet{heyer-turb} are
developed to measure the structure function of turbulence within
molecular gas and provide an excellent descriptor of the turbulence.
The dendrogram application provides a similar measurement by measuring
the spatial and velocity extent of the isosurfaces within an emission
data cube.  Indeed, the virial analysis above can be thought of
comparing the spatial and velocity extents of the isosurfaces to the
amount of emission they contain.  We can construct the size-line width
relationship within a data cube by plotting the size vs. the line
width of the isosurfaces in the data.  This becomes, in effect, a
``Type 4'' size-line width relationships discussed in
\citet{goodman-coherence}; that is to say, using a single tracer
species to measure the relationship in a single cloud.  

\begin{figure}
\plotone{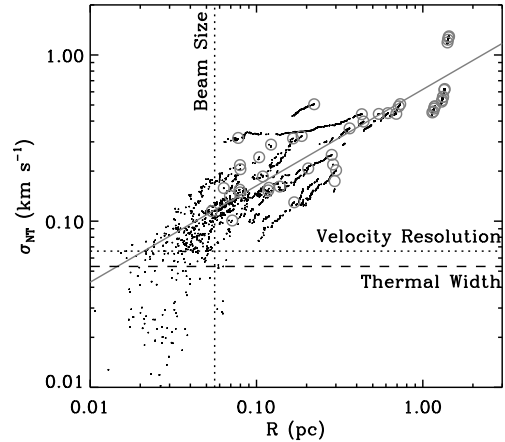}
\caption{\label{rdv} The size-line width relationship for the
  isosurfaces of $^{13}$CO emission in L1448.  The thermal line width
  (for $T=10$~K) and the limits of the instrumental resolution are
  also indicated in the plot and data below these limits are
  unreliable.  Individual points correspond to isosurfaces used in our
  contouring.  Gray circles indicate the characteristic size and line
  width for each {\it independent} branch in the dendrogram and
  represent a useful minimum sampling of the data. The gray line
  indicates the fit to the gray, circled data: $\sigma_v = 0.6
  R^{0.58}$ which is comparable to the size-line width relationship
  found among turbulent molecular clouds.}
\end{figure}

In Figure \ref{rdv}, we plot the size-line width relationship for the
$^{13}$CO emission from L1448.  A fit to the grey, circled data gives
$\sigma_{\mathrm{NT}}= (0.62\pm 0.04) R_{\mathrm{pc}}^{0.58\pm
  0.04}\mbox{ km s}^{-1}$ typical of the turbulent gas in molecular clouds
\citep{goodman-coherence}.  The scaling relations for turbulent gas
traces the data at large scales/line width well.  The size has been
corrected for beam convolution effects by subtracting the beam width
in quadrature.  Similarly, the line width has had the thermal
contribution of 10 K gas removed.  These corrections are rough since a
simple Gaussian deconvolution is inappropriate for the small
isosurfaces defined by high brightness contours and the broadening
effect of the spectrometer has been neglected.  Errors arising from
our approximate treatment will become relatively small for sizes and
line widths that are significantly larger than the instrumental
resolution.  At large line widths, linear sets of points correspond to
branches in the dendrogram.  Along branches, the properties change
slowly (see, for example, the radius and luminosity in Figure
\ref{propplot}).  The discontinuities occur when two isosurfaces
merge together resulting in an abrupt change in the size and the line
width of the isosurface.  At small scales, the data dissolving into a
sea of noise since the properties of these surfaces are poorly defined
in the face of thermal noise and instrumental convolution effects.
The scatter around the average relationship results, in part, from the
details of the isosurface shapes which vary due to turbulent
velocity/density fluctuations but also due to noise. 

Dendrograms can also be used to abstract the data set to a defining
set of isosurfaces.  The gray circles shown in Figure \ref{rdv} plot a
single, characteristic size and line width for each branch of the
dendrogram shown in Figure \ref{l1448color}.  Since the dendrogram
contains 26 leaves, there are 51 independent branches ($2N-1$) because
each branch is required to join with others in a binary merger.
Hence, only one point is plotted for every significantly distinct set
of isosurfaces in the data and multiple points from redundant
isosurfaces are suppressed. 

\section{Identifying Giant Molecular Clouds}
\label{gmcs}

An additional application of the dendrogram technique is to identify
Giant Molecular Clouds (GMCs) in a blended data set.  Massive,
isolated molecular clouds show virial parameters close to unity
\citep{srby87,hc01}.  We propose that GMCs can be identified as
the largest-scale self-gravitating structures in the ISM and such
structures can be identified in the dendrogram analysis.  Unlike the
substructure analysis presented previously, we are only interested in
the dynamical state of the largest scale emission and the
contamination by background emission is likely minimal.  Hence, the
property calculations for the dendrogram can use the extrapolation
paradigm since we are interested in the properties at the 0 K
km~s$^{-1}$ isosurface.

To demonstrate this application of the dendrogram technique, we use
the Orion-Monoceros $^{12}$CO data of \citet{orion-wil} taken with the
CfA 1.2-m telescope.  All of the Orion-Monoceros complex is contained
within a single isosurface with $T_{mb} =0.4$~K which must be
decomposed into the constituent GMCs.  We adopt the standard
CO-to-H$_2$ conversion factor ($X_2=1$) and use the extrapolation
paradigm to calculate the virial parameter for each branch of the
dendrogram.  We adopt a distance to the main Orion complex of 450 pc
and use a distance of 800 pc for Monoceros, 425 pc for NGC 2149, and
400 pc for the Northern Filament based on the identifications and
distance estimates of \citet{orion-wil}.

\begin{figure*}
\plotone{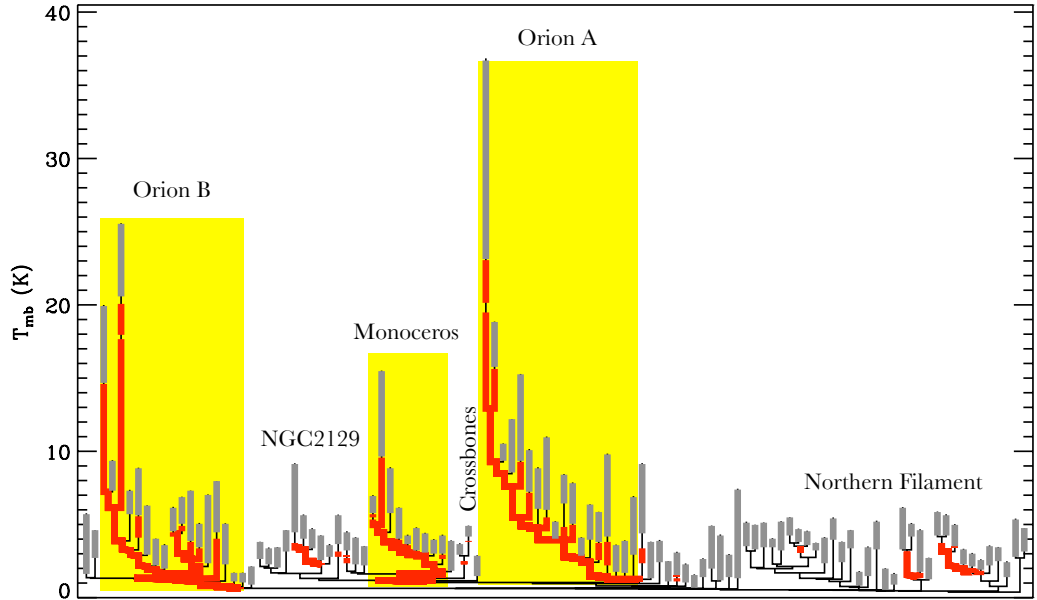}
\caption{\label{orionfig} The dendrogram of the Orion-Monoceros
  region.  Branches of the dendrogram corresponding to
  self-gravitating structures are highlighted in red.  Regions where
  the quality of the data prohibit accurate estimation of the virial
  parameter are shown in gray.  The GMCs within the data cube are
  identified as the largest scale objects that are self-gravitating
  but not bound to each other.  Regions of the dendrogram
  corresponding to specific objects are labeled and the sections of
  the dendrogram corresponding to GMCs are shaded in yellow.}
\end{figure*}
\begin{figure}
\plotone{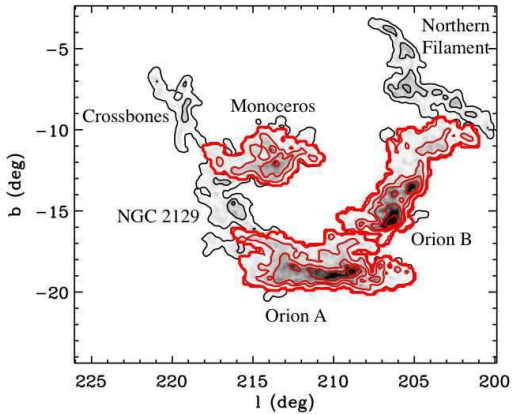}
\caption{\label{orionmap} Map of emission for the Orion-Monoceros
  region contained within a $T_{mb}=0.4$~K contour.  The three
  constituent GMCs in the complex have been identified using the
  dendrogram analysis and their boundaries are indicated in red.  The
  regions are labeled according to their designations in
  \citet{orion-wil}.}
\end{figure}

Figure \ref{orionfig} shows the dendrogram for the region with sets of
isosurfaces that have $\alpha <2$ highlighted.  We identify GMCs in an
automatic fashion as all emission contained within distinct,
self-gravitating regions with masses $M>5\times 10^4~M_{\odot}$.  The
three GMCs in the data cube naturally segregate from the rest of the
emission.  In Figure \ref{orionmap} we show the emission contained in
the $T_{mb}=0.4$~K contour and its characteristic designations.  The
dendrogram analysis identifies three regions as GMCs and finds that
the remaining emission is not sufficiently massive or self-gravitating
to be identified as a GMC.  Given the good agreement with the standard
identifications, we conclude that the dendrogram method can be used to
identify GMCs in blended sets of emission.  The primary restriction is
good knowledge of the distances to different regions of the data
volume.  This limitation implies that the method is best applied in
the outer galaxy or in extragalactic analysis where distances are
well-determined.

\section{Summary}

We have presented a new application of tree diagrams to
three-dimensional data sets.  This application is closely related to
the structure tree analysis of \citet{houlahan2}.  These techniques
use dendrograms to represent the merging/bifurcating of contours in a
three-dimensional data set as a function of contour level.  Each point
in the dendrogram corresponds to an isosurface in the data cube.  By
characterizing molecular emission associated with these isosurfaces,
we are able to measure the properties of both small- and large-scale
structures in a data set.  In particular, we emphasize determinations
of the virial parameter and the size-line width relationship at
multiple scales in the data.  The virial parameter, in particular,
provides a means to estimate the influence of self-gravity on a
variety of scales in the molecular cloud yielding, for the first time,
a uniform study of energetics on a range of scales.

The dendrogram technique is philosophically different from
segmentation algorithms such as CLUMPFIND.  By preserving and
characterizing the hierarchy of emission isosurfaces in a data cube,
it is possible to study structures over a range of scales.  In
principle, the results of the hierarchical decomposition is independent
of algorithm parameters, though the actual output is governed by the
degree of simplification desired by the user.  Although the dendrogram
analysis does not segment the data by itself, the results can be used
to provide a physically-motivated segmentation of objects in some
systems.  

Despite the power of the dendrogram technique there are worrisome
ambiguities at relating the observed to the physical domain.  We have
presented three attempts to account for ambiguities in this
relationship, but find no satisfactory, universally-applicable
method.  We note that many of our results are subject to caveats
regarding interpretations, but we argue that these caveats do not
undermine the applicability of the techniques.

We have analyzed $^{13}$CO($1\to 0$) emission from the L1448 region as
observed by the COMPLETE survey of Perseus \citep{complete-data} using
the dendrogram methods.  We note that common structure analysis
techniques have a fundamental scale built into their analysis and tend
to analyze the dynamical state of objects on that scale.  As such, the
synthesis of many such analytic studies conducted on various scales
leaves an impression that the dendrogram analysis actually
demonstrates.  We find self-gravitating structures on all scales in
L1448, though not in all regions.  In particular, the majority of
emission in small-scale structures is {\it not} self-gravitating; but,
at larger scales, much of the L1448 region is influenced by self
gravity.

We have also illustrated the capacity for the dendrogram technique to
make differential measurements between data sets.  The dendrogram of
L1448 is compared to a dendrogram of a theoretical simulation finding
qualitative and quantitative differences.  Differences of this
magnitude were not discernible through other statistical techniques
such as determinations of the power spectra \citep{padoan-perseus}.
Future work will investigate further applications of the differential
measurement techniques between dendrograms.

The dendrogram technique can be used to measure the size-line width
relationship within molecular clouds using the characteristic sizes
and line widths of the constituent isosurfaces in the data.  As
expected, we recover the typical size-line width relationship for
molecular clouds $\sigma_v \propto R^{0.58}$ within a single cloud.  

Finally, we conclude the paper by presenting an alternative
application of dendrograms: the identification of Giant Molecular
Clouds in blended line data sets.  We define GMCs as massive
($M>5\times 10^4$) clouds of gas that are (a) self-gravitating but (b)
not bound to their surrounding medium.  This definition not only
identifies GMCs but does so exclusive from including low-mass chaff
that is dynamically unrelated to the GMCs.  Using this simple
definition the dendrogram technique readily identifies the three
constituent GMCs in the blended Orion-Monoceros data of
\citet{orion-wil}.  

Beginning with common application of techniques developed previously,
this new perspective on dendrograms illustrates their utility at the
visualization and reduction of molecular line data.  Dendrograms
reduce three dimensional hierarchical data sets to a two dimensional
plot that retains essential features regarding the topology of the
emission.  This reduction is conducted in a fashion that is minimally
model dependent, relying on the intrinsic structure of the
isosurfaces in an emission line data set.

\acknowledgements 

We are grateful for useful discussions with many people concerning the
application, development, and visualization of dendrograms.  In
particular, we thank Mike Halle, Michelle Borkin, Jonathan Foster,
Jonathan Williams, Paola Caselli, and Mark Heyer for constructive
comments regarding the development of dendrograms.  The comments of an
anonymous referee improved the presentation and accuracy of this work.
We thank Tom Dame for the use of the Orion-Monoceros data cube.  ER's
work is funded by an NSF Astronomy and Astrophysics Postdoctoral
Fellowship (AST-0502605).  JEP is supported by the National Science
Foundation through grant \#AF002 from the Association of Universities
for Research in Astronomy, Inc., under NSF cooperative agreement
AST-9613615 and by Fundaci\'on Andes under project No. C-13442. This
material is based upon work supported by the National Science
Foundation under Grant No. AST-0407172. This work made extensive use
of the NASA's Astrophysics Data System.


\end{document}